\documentclass[11pt,twoside]{article}
\usepackage{hepro5}
\usepackage{graphicx}

\usepackage[T1]{fontenc} 

\usepackage{latexsym}
\usepackage{verbatim}

\begin{document}

\vskip 1.0cm
\markboth{Tavecchio \& Bonnoli}{Extreme BL Lacs}
\pagestyle{myheadings}

\vspace*{0.5cm}
\title{Extreme BL Lacs: probes for cosmology and UHECR candidates}

\author{F. Tavecchio and G. Bonnoli}
\affil{INAF-OAB, Via Bianchi 47, 23807, Merate, Italy}

\begin{abstract}
High-energy observations of extreme BL Lac objects, such as 1ES0229+200 or 1ES 0347--121, recently focused interest both for blazar and jet physics and for the implication on the extragalactic background light and intergalactic magnetic field estimate. Moreover, their enigmatic properties have been interpreted in a scenario in which their primary high- energy output is through a beam of high-energy hadrons. However, despite their possible important role in all these topics, the number of these extreme highly peaked BL Lac objects (EHBL) is still rather small. Aiming at increase their number, we selected a group of EHBL candidates considering those undetected (or only barely detected) by the LAT onboard Fermi and characterized by a high X-ray versus radio flux ratio. We assembled the multi-wavelength spectral energy distribution of the resulting 9 sources, using available archival data of {\it Swift}, {\it GALEX}, and {\it Fermi} satellites, confirming their nature. Through a simple one-zone synchrotron self-Compton model we estimate the expected very high energy flux, finding that in the majority of cases it is within the reach of present generation of Cherenkov arrays or of the forthcoming CTA.
\end{abstract}

\section{Introduction}

Intense emission of gamma rays is a distinctive feature of blazars, active galactic nuclei (AGN) dominated by the relativistically boosted non--thermal continuum from a relativistic (typical bulk Lorentz factors $\Gamma=10-20$) jet pointing toward the observer. The spectral energy distribution (SED) of blazars displays two  broad bumps, whose peak frequency appears to anti-correlate with the emitted luminosity -- the so-called {\it blazar sequence} (Fossati et al. 1998, but see Giommi et al. 2005). The low energy (from radio  to optical-UV bands) emission is clearly associated to synchrotron radiation from relativistic electrons (or pairs). The nature of the mechanisms responsible for the high-energy component, instead, is still debated. The most popular view adopts the so-called leptonic scenario (e.g. Ghisellini et al. 1998), in which the high energy radiation is interpreted as inverse-Compton emission by the same leptons responsible for the synchrotron component. Alternatively, hadronic models interpret the high-energy emission as either the by-product of cascades initiated by ultra-relativistic hadrons  or synchrotron emission from high-energy protons (e.g. Muecke et al. 2003). 

A quite interesting feature of blazars is the intense very high energy (VHE, $E>100$ GeV) emission, characterizing in particular the subclass of BL Lac objects. The great majority of known TeV BL Lac objects have the maximum of the the high-energy component peaking in the 1-100 GeV band. Therefore, typical VHE spectra 
of BL Lacs are soft. However, there is a (still) small group of TeV BL Lacs  for which the maximum is located well above 1 TeV (e.g. Tavecchio et al. 2011). These so-called {\it extreme} HBL (Costamante et al. 2001) represent a challenging case for the standard leptonic scenario, apparently requiring a quite special physical set-up (e.g. Katarzy{\'n}ski et al. 2005, Tavecchio et al. 2009, Lefa et al. 2011). Besides the importance for blazar emission models, the extremely hard gamma-ray emission offer the ideal tool to probe absorption by the intervening extragalactic background light (EBL, e.g., Costamante 2013 and references therein), or to measure the tiny intergalactic magnetic field permeating the extragalactic space (e.g. Tavecchio et al. 2010, Neronov \& Vovk 2010). In this context, EHBL offer also an useful tool to probe possible deviations of the gamma-ray propagation, related either to the mixing of photons with new particles (axion-like particles, e.g. De Angelis et al. 2011) or to the violation of the Lorentz invariance et high energy (e.g. Tavecchio \& Bonnoli 2015).

Leptonic models do not naturally explain the quite limited variability of the VHE emission of EHBL, at odds with the extreme (both in amplitude and in time-scale) variability characterizing the high-energy emission of the average BL Lac population. This evidence is instead naturally reproduced in an alternative view, postulating that the observed very high energy (VHE) emission is the left-over of electromagnetic cascades, occurring in the intergalactic space, possibly triggered by a beam of high-energy hadrons produced in the jet of the EHBLs. In fact, high-energy hadrons (proton for simplicity) accelerated and beamed by the jet would interact with low energy background photons, producing neutral pions that, in turn, decay in ultra-high energy photons, which initiate a cascade. In this framework, misaligned jets of EHBL residing within the GZK horizon could contribute to the flux of ultra-high energy cosmic rays at the Earth (Tavecchio 2014). A nice testable prediction of the model sketched above is that the existence of an observable hard tail above 10 TeV (energies at which, instead, due to the severe absorption leptonic models do not predict any detectable signal), see e.g. Murase et al. (2012).

The known EHBL are still quite a few. One reason for this situation is the strong bias of {\it Fermi}-LAT against the detection of EHBL which in the GeV band are characterized by a quite weak emission. With the aim of enlarging the sample of these extremely interesting sources, we started a program of selection of candidates EHBL based on the peculiar properties of the multifrequency SED of these sources. In the following we outline the selection procedure and the results. For more details we refer the reader to the paper by Bonnoli et al. (2015).

\section{Enlarging the EHBL population}

A distinctive property of the SED of EHBL is to display a low radio flux together with a relatively bright X-ray emission. This feature provides an affective way to separate EHBL from the population of ``normal" BL Lac objects. We exploit this strategy starting from  a list of 71 BL Lac (Plotkin et al. 2011) resulting from the correlation of SDSS and FIRST surveys and by optical spectrum dominated by the host galaxy emission. We further select the sources with measured X-ray flux (by {\it ROSAT}) and we restrict the redshift to $z< 0.4$, so that the absorption of TeV gamma rays is not too large, obtaining a total of 50 BL Lacs. 

\begin{figure}  
\begin{center}
\hspace{0.25cm}
\includegraphics[height=10.0cm]{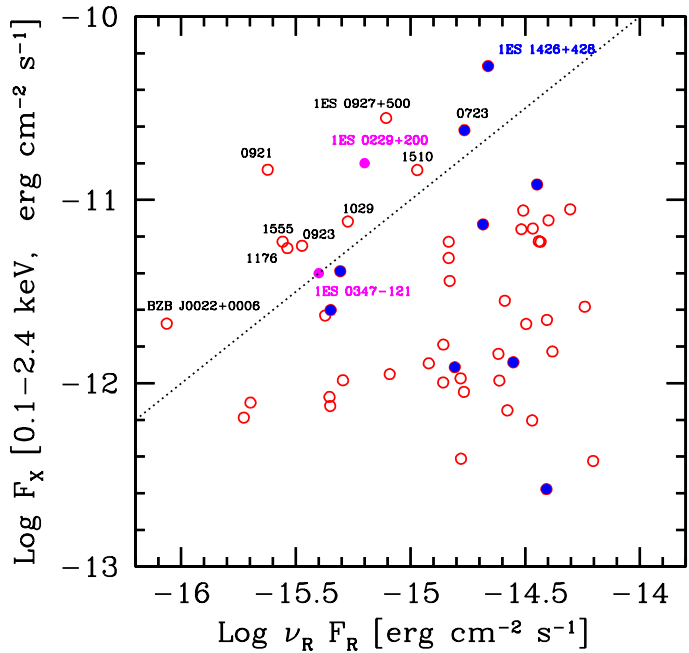}
\caption{F$_X$--F$_R$  plot for the sample of BL Lacs described in the text. Red open circles represent the full sample; the blue circles are the sources detected by LAT in the 1FGL catalogue. The 12 sources above the  black dotted diagonal line
  have a high ($F_X/F_R > 10^4$) ratio of X--ray vs. radio  flux. The magenta filled circles are the
  two archetypal TeV detected, but GeV faint extreme BL Lacs, 1ES 0229+200 and 1ES 0347-121. From Bonnoli et al. (2015).}
\label{frfx}
\end{center}
\end{figure}

The X-ray {\it vs.} the radio flux of the sources are shown in Fig. \ref{frfx}. According to our criterion, we select the EHBL candidates among the sources residing in the left upper corner of the plot (low radio flux, large X-ray flux), above the dotted line. This choice is consistently supported by the fact that  two prototypical EHBL (1ES 0229+200 and 1ES 0347-121) appear in this region (magenta points). Eliminating the already known EHBL, we obtained a total of 9 new candidates, whose nature is further investigated using the SED.

We assembled the SED exploiting archival data (especially by {\it Swift} and {\it GALEX}) and asking dedicated {\it Swift} pointings for RBS 0923 and RBS 1176, for which previous observations were not available. {\it Swift} observations are crucial for characterizing  these sources, since UVOT and XRT data allow us to outline the UV-X-ray continuum, a key band for the identification of EHBL. Indeed, the shape of the low-energy (i.e. synchrotron) SED component of all the sources confirm the EHBL nature of all selected candidates. In particular, for most of the sources the {\it Swift}/UVOT and {\it GALEX} data track the UV  tail of the host galaxy, sometimes showing a hint of the expected hard jet emission at the highest frequencies. As expected, the slope of the X-ray spectra (which, we remark, was not used in the selection procedure) are on average  flat (photon index around 2), indicating that the peak of the synchrotron component occurs in this band.

\subsection{Some interesting cases}

\begin{figure}  
\begin{center}
\hspace*{-.5cm}
\includegraphics[height=14.5cm,width=15.0cm]{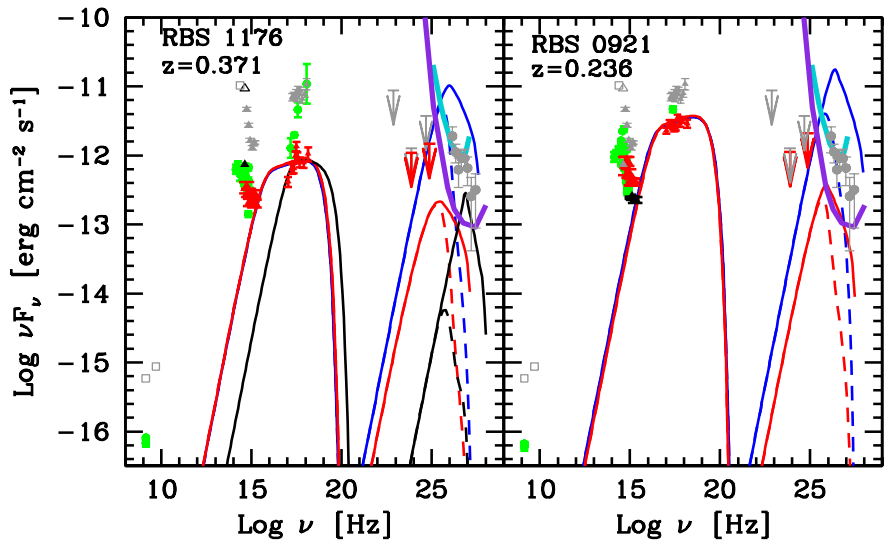}
\vspace*{-3.5 truecm}
\caption{SED of two EHBL candidates selected in Bonnoli et al. (2015). Green symbols report historical data, red symbols show {\it Swift}/UVOT, XRT and {\it Fermi}/LAT data, black symbols display  {\it GALEX} data. Background grey symbols show the SED of the prototypical EHBL 1ES 0229+200 for comparison.
   Two different models of the SED are reported, corresponding to low ($B=0.01$ G,
  blue lines) and high ($B=0.1$ G, red lines) magnetic field.  Dashed lines show the model after
  absorption with the EBL.  Light blue and violet curves report the differential sensitivities ($5\, \sigma$, 50
  hours of exposure, 5 bins per energy decade) of MAGIC and CTA respectively.}
\label{sed}
\end{center}
\end{figure}

In Fig. \ref{sed} we show the SED of two particularly interesting sources, together with a prediction of their VHE emission based on the one-zone synchrotron-self Compton model. We report two cases, corresponding to two values of the magnetic field ($B=0.1$ and 0.01 G, red and blue lines respectively), expected to bracket the range valid for these sources. Modelling the synchrotron component, tracked by the UV-X-ray data, we derive the electron distribution which, in turn, allows us to  specify the SSC component. The intrinsic SSC emission (solid line) is then corrected to take into account the absorption of gamma-rays through the interaction with the EBL (dashed lines).

RBS 1176 (left panel), displays a peculiar X-ray spectrum, with a rather hard slope at low energies. While it cannot be excluded that the lack of soft photons is related to absorption (intervening or internal), it is tempting to interpret the spectral break as directly linked to the electron distribution. In fact, the break can be well reproduced by assuming that the electrons follow a power law in energy, with a very large minimum energy $E_{\rm min}$, so that the hard low-energy X-ray continuum is reproduced by the $\nu^{1/3}$ low-energy tail of the synchrotron emission of electrons with $E_{\rm min}$. In this case (black line in Fig. 2), the predicted SSC component would peak at energies well above 10 TeV, but the predicted flux would be vary low, due to the very small flux of the target low-energy synchrotron photons. The comparison with the CTA sensitivity curve (calculated for 50 hours of exposure and 5$\sigma$ significance), shows that, in this case, the detection at VHE would be unlikely, also because the relatively large redshift causes a quite pronounced drop of the observed VHE flux.

The right panel of Fig. \ref{sed} instead shows a quite promising source, RBS~0921.  The bright and hard X-ray continuum (indicating that the peak of the synchrotron component occurs at energies larger then several keV) translates into a relatively bright VHE component, possibly detectable even with current Cherenkov arrays such as MAGIC (light blue curve).

\section{Discussion and prospects}

A generalization of our criterion, for instance relaxing the condition of the redshift, will allow us to extend the selection procedure to larger samples of BL Lacs (or, possibly, radio-loud AGN). A larger number of EHBL  is especially important for improving our understanding about the far-IR EBL, IGMF and the extragalactic gamma-ray background. Exploiting the capability of {\it NuStar}, SKA, and CTA will also be particularly revealing.  Even before the construction of the full CTA, the planned ASTRI/CTA  mini--array (Vercellone et al. 2013) could be used in this direction. It is arguable that HAWC  and later on LHAASO  will effectively survey them in the local Universe.

An interesting topic to be addressed is the position of EHBLs within the general BL Lac population and, especially, the nature of their parent population. In the framework of the standard one-zone leptonic model the physical conditions within the EHBL jets appear to be different from that of the other EHBL, with rather low magnetic fields and large particle energies. It is not clear whether and how these differences are related to the other striking difference of EHBL, i.e. the very limited variability. The differences of EHBL could be somewhat related to the different parent population. By construction, the parent population are expected to have very faint radio emission (radio luminosities around $L_r\sim 10^{40}$ erg  s$^{-1}$) indicating very weak radio jets (see Giroletti et al. 2004). Considering that at least a fraction of the radio flux comes from the beamed jet synchrotron component, we argue that the intrinsic radio luminosity drops below the lower end of the FRI radiogalaxy power range, making  EHBLs suitable candidates for the aligned counterparts of the weak radiogalaxies population (dubbed FR 0) studied by Baldi \& Capetti (2010), Baldi et al. (2015).

\acknowledgments We acknowledge financial support by PRIN-INAF 2014 and by the CaRiPLo Foundation and the Regional Government of Lombardia to the project ID 2014-1980 ``Science and technology at the frontiers of gamma-ray astronomy with imaging atmospheric Cherenkov Telescopes"

\end{document}